\documentclass[aps,reprint,prl,showpacs]{revtex4-2}

\usepackage{cancel}
\usepackage[pdftex,bookmarks,colorlinks,citecolor=blue,linkcolor=blue]{hyperref}
\usepackage{newtxtext}
\usepackage[varg]{newtxmath}
\usepackage{bm}
\usepackage{graphicx}
\usepackage{url}
\usepackage{xcolor}
\usepackage{ulem}

\newcommand{\D}{\mathrm{d}}

\newcommand\op[3]{\bm{#1}^{#2}_{#3}}

\newcommand\R[1]{\op{r}{}{#1}}

\newcommand{\gd}{\dot{\gamma}}

\newcommand{\DUinf}{\bm{\nabla}\bm{u}^{\infty}}
\newcommand{\Einf}{\bm{E}^{\infty}}
\newcommand{\Ominf}{\bm{\Omega}^{\infty}}

\newcommand{\giso}{g_{\rm iso}}
\newcommand{\gSigma}{{\bm \Sigma}}

\newcommand{\gQ}{\bm{Q}}

\newcommand{\gF}{\bm{f}}

\newcommand\Msymt[2]{\begin{pmatrix} #1 & #2 \\ #2 & -#1 \end{pmatrix}}

\newcommand\gdot{\bm{\cdot}}

\newcommand{\er}{\bm{e}_r}

\newcommand{\eroer}{\er\otimes\er}

\newcommand{\rs}{{r^*}}

\def\identity{\;\mbox{l\hspace{-0.55em}1}}

\begin{document}

\title{Microscopic theory for the rheology of jammed soft suspensions}

\author{Nicolas Cuny}

\author{Romain Mari}

\author{Eric Bertin}
\affiliation{Univ. Grenoble Alpes, CNRS, LIPhy, 38000 Grenoble, France}

\date{\today}

\begin{abstract}
We develop a constitutive model allowing for the description of the rheology of two-dimensional soft dense suspensions above jamming. Starting from a statistical description of the particle dynamics, we derive, using a set of approximations, a non-linear tensorial evolution equation linking the deviatoric part of the stress tensor to the strain-rate and vorticity tensors. The coefficients appearing in this equation can be expressed in terms of the packing fraction and of particle-level parameters.
This constitutive equation rooted in the microscopic dynamic qualitatively reproduces a number of salient features of the rheology of jammed soft suspensions, including the presence of yield stresses for the shear component of the stress and for the normal stress difference. More complex protocols like the relaxation after a preshear are also considered, showing a smaller stress after relaxation for a stronger preshear.
\end{abstract}

\maketitle

Soft athermal suspensions are mixtures of non-Brownian soft elastic units in a fluid, like microgels or emulsions~\cite{bonnecaze_micromechanics_2010}.
They are commonly used for their ability to turn from soft elastic solids to liquids under applied stress when the particle volume fraction $\phi$ is large enough, 
e.g.~mayonnaise~\cite{ma_rheological_1995} or hand sanitizer gels~\cite{roberts_new_2001}.
Indeed, above a jamming concentration $\phi_\mathrm{J}$, a finite yield stress $\sigma_\mathrm{y}$ develops~\cite{liu_jamming_1998}.
The steady state rheology under simple shear at a rate $\dot\gamma$ follows a Herschel–Bulkley (HB) law $\sigma = \sigma_\mathrm{y} + k \dot\gamma^n$, with $n\leq 1$~\cite{piau_carbopol_2007,moller_attempt_2009,divoux_transient_2010,seth_micromechanical_2011,liu_universality_2018}, which extends to more general deformations~\cite{balmforth_yielding_2014}. 
The HB rheology for $\phi \geq \phi_\mathrm{J}$ is also well documented from numerical simulations 
of generic minimal particle models~\cite{otsuki_universal_2009,otsuki_critical_2011,olsson_herschel-bulkley_2012,kawasaki_diverging_2015}, as well as more system-specific models~\cite{seth_micromechanical_2011,gross_rheology_2014}. 

The transient mechanical response is also rich. 
If a sample initially at rest is brought to yield, the load curve often shows a stress overshoot, 
that is, a local maximum of stress before a further decay to its steady state value, for deformations of order 1~\cite{partal_transient_1999,batista_colored_2006,divoux_stress_2011,kaneda_stress_2011,younes_elusive_2020}.
By contrast, when a flowing sample is brought to rest by suddenly stopping the applied deformation, the stress relaxes below the yield stress with a counter-intuitive dependence on the previous deformation: the larger the stress during the initial shear, the smaller the residual stress~\cite{mohan_microscopic_2013,mohan_build-up_2014,lidon_power-law_2017}.

Most current theoretical approaches address yield stress fluids (YSF) in general, that is, a much wider class of systems 
including soft jammed suspensions but also gels or colloidal glasses~\cite{bonn_yield_2017}.
They are oblivious to the microscopic origin of the yield stress, which they take for granted at a coarser level.
Continuum models postulate evolution equations for the macroscopic stress tensor assuming the presence of a yield stress.
For YSF, they fall into two main classes, viscoplastic models if they ignore the elasticity below yield~\cite{papanastasiou_flows_1987} or elastoviscoplastic models if they consider it~\cite{saramito_new_2007,saramito_new_2009,park_oscillatory_2010,belblidia_computations_2011}.
While successful at describing even complex flow situations~\cite{cheddadi_understanding_2011}, 
their structure and parameters are usually 
not explicitly connected to microscopic properties.
Mesoscopic models, such as Soft Glassy Rheology~\cite{sollich_rheology_1997,sollich_rheological_1998}, elasto-plastic models~\cite{nicolas_deformation_2018} or Shear Transformation Zone (STZ) theory~\cite{falk_dynamics_1998,bouchbinder_athermal_2007}
explain the emergence of the nontrivial YSF rheology by the statistics of large assemblies of simple plastic mesoscopic units, possibly mechanically coupled. 
Nonetheless, plasticity is assumed at the mesoscopic level, and comes in many variants~\cite{hebraud_mode-coupling_1998,bouchbinder_athermal_2007,nicolas_rheology_2014,lin_scaling_2014}.
Finally, Mode-Coupling Theory (MCT) is the only microscopic theory addressing YSF (beyond simple scaling arguments~\cite{tigheModelScalingStresses2010}), 
but it focuses on thermal colloidal glasses, for which the yield stress arises 
from the glass rather than the jamming transition~\cite{fuchs_theory_2002,brader_glass_2009}.

In this Letter, we aim at developing a rheological model for jammed athermal soft suspensions from their microscopic dynamics.
The goal is to get a temporal evolution for the stress tensor taking the deformation rate tensor as input, which structure and parameters can be directly related 
to particle properties.
We follow a microstructure route based on a dynamical equation for the pair correlation function.
While this approach is common for near equilibrium polymeric systems~\cite{doi_theory_1988},  it is much less explored for non-Brownian suspensions, which are athermal and thus far from equilibrium states.
Previous works in this direction addressed colloidal suspensions below jamming~\cite{lionberger_smoluchowski_1997,nazockdast_microstructural_2012,banetta_pair_2020}, 
a regime dominated by Brownian motion and hydrodynamics.
Our work focuses on jammed suspensions, for which the main source of stress is the elastic deformation of particles~\cite{liu_universality_2018}.

We consider a model of a two-dimensional suspension consisting of $N$ identical soft discs of radius $a$ immersed in a viscous fluid, akin to the Durian bubble model~\cite{durian_foam_1995}.
Particles have an overdamped, non-Brownian dynamics, and interact through radial contact repulsion forces only. 
The externally applied deformation generates a velocity field $\bm{u}^{\infty}(\bm{r})$ in the fluid, 
which for the sake of simplicity we assume affine, $\bm{u}^{\infty}(\bm{r}) = \nabla\bm{u}^{\infty} \cdot \bm{r}$
(we define the velocity gradient as $(\nabla\bm{u}^{\infty})_{ij}=\partial \bm{u}^{\infty}_i/\partial r_j$).
We consider the influence of the fluid through a viscous drag
$-\lambda_{\rm f}\big(\dot{\bm{r}}_{\mu}-\op{u}{\infty}{}( \op{r}{}{\mu})\big)$, neglecting hydrodynamic interactions between particles, an assumption justified by screening effects in  dense systems.
The position $\bm{r}_{\mu}$ of particle $\mu =1,\dots,N$ evolves according to
\begin{equation}
\label{eq_u}
\dot{\bm{r}}_{\mu} = \op{u}{\infty}{}( \op{r}{}{\mu}) + \frac{1}{\lambda_{\rm f} }\sum\limits_{\nu=1}^{N}
\gF(\R{\nu}{}-\R{\mu}{})\,.
\end{equation}
We consider a radial contact force $\gF(\R{})$, $\gF(\R{})=f(r)\, \er$, with $r=||\R{}||$ and $\er=\R{}/r$.
Note that $f(r)<0$ for a repulsive force.

We now outline our coarse-graining procedure, whose details are reported elsewhere~\cite{CunyPRE21}. We start from the exact time evolution for the pair correlation function $g(\R{})$ obtained from the conservation of the $N$-particle probability, 
\begin{equation}
\begin{split}
\label{eq_g2}
\partial_t g(\R{}) + \bm{\nabla} \gdot \Big[ & \left(\DUinf \gdot \R{}\right) g(\R{}) -\gF(\R{})g(\R{})  \\ 
&-\rho \int \gF(\bm{r}') g_3(\R{},\bm{r}') \D \bm{r}'\Big] = 0 \,,
\end{split}    
\end{equation}
using the particle radius $a$ as unit length, a characteristic force $f_0$ as unit force and $\tau_0=\lambda_{\rm f} a/(2f_0)$ as unit time.
Eq.~\eqref{eq_g2} is not closed, but involves  the three-body correlation function
$g_3(\bm{r},\bm{r}')$  whose evolution equation also involves higher order correlation functions, in a hierarchical manner.

The particle stress tensor $\gSigma$ is defined from the Virial formula \cite{nicot_definition_2013} as
\begin{equation}
	\label{sigma_int}
		\gSigma = \frac{\rho^2}{2} \int \big(\R{} \otimes \gF(\R{}) \big) \, g(\R{}) \, \D\R{}.
\end{equation}
Multiplying Eq.~\eqref{eq_g2} by $\frac{1}{2}\rho^2 \, \R{} \otimes \gF(\R{})$ and integrating over $\R{}$, we get the following evolution equation for  $\gSigma$,
\begin{equation}
	\label{eq:Sigma:v0}
	\dot{\gSigma} =\nabla \op{u}{\infty}{} \gdot \gSigma + \gSigma \gdot \nabla \op{u}{\infty\,T}{} + \bm{G}_2 - \bm{G}_3
\end{equation}
where $\bm{G}_2$ and $\bm{G}_3$ are defined as
\begin{align}
	\label{def_G2}
	\bm{G}_2 &= \frac{\rho^2}{2} \int \left[ \big(\Einf:\eroer\big)
	\left( (\R{}\otimes\R{})\cdot\nabla \gF(\R{})-\R{}\otimes\gF(\R{})\right)\right.\nonumber\\
		& \qquad \left. -\gF(\R{})\otimes\gF(\R{}) -\big(\R{}\otimes\gF(\R{})\big)\cdot\nabla \gF(\R{})^{T} \right] g(\R{}) \D\R{}\,, \\
	\label{def_G3}
	\bm{G}_3 &=\frac{\rho^3}{2} \iint \left[\gF(\bm{r}')\otimes \gF(\R{}) +
	\left(\R{} \otimes \gF(\bm{r}')\right)\gdot \nabla\gF(\R{})^{T}\right] \nonumber \\ & \qquad \qquad \qquad \qquad \qquad \qquad \quad \times g_3(\R{},\bm{r}')\D\R{}\D\bm{r}'\,.
\end{align}
To find a closed equation on $\gSigma$, 
we use the Kirkwood closure~\cite{kirkwood_statistical_1935}, which approximates $g_3$ as a product of pair correlation functions,
\begin{equation}
	\label{kirkwood}
	g_3(\R{},\bm{r}')=g(\R{})g(\bm{r}')g(\R{}-\bm{r}'),
\end{equation}
which is a reasonable approximation for interparticles distances relatively close to the particle diameter (i.e., particles close to contact) \cite{CunyPRE21}.
Then, we parameterize $g(\bm{r})$ in terms of $\gSigma$ to close Eq.~\eqref{eq:Sigma:v0}.
In the isotropic state, the function $g(\R{})$ is peaked on a circle in the $\R{}$-plane, whose radius $r^*$ maximizes
$\giso(r)$. In the presence of a weak anisotropy,
one expects the locus of the maxima of $g(\R{})$ to deform 
according to a small amplitude second order harmonic, which leads us to approximate $g(\R{})$ as a small deformation of $\giso(r)$ of the form
\begin{equation}
	\label{approx_g}
	g(\R{}) = \giso\left(\frac{r}{1-\alpha\left(\gQ:\eroer\right)}\right),
\end{equation}
with $\alpha$ a parameter to be determined,
and where the traceless structure tensor $\gQ$ (an anisotropy measure) is defined as
\begin{equation}
\label{eq:def:Q}
\gQ=\frac{\rho^2}{2}\int_{r\leq 2} \left( \R{}\otimes\R{} -\frac{r^2}{2}\identity \right) g(\R{})\D\R{}.
\end{equation}
Expanding Eq.~\eqref{approx_g} to first order in $\gQ$, one finds
\begin{equation}
	\label{DL_g}
		g(\R{}) \approx \giso(r) + \alpha r \giso'(r) \left(\gQ:\eroer\right).
\end{equation}
Self-consistency between \eqref{eq:def:Q} and \eqref{DL_g} imposes that $\alpha \int_0^2 r^4 \giso'(r) \D r = \frac{4}{\pi\rho^2}$.
Using the definition \eqref{sigma_int} of $\gSigma$ and the parametrization \eqref{DL_g} of $g(\R{})$, we get the deviatoric stress
$\gSigma' = k \gQ$,
with $k = \frac{\pi\alpha \rho^2}{4}\int_0^2 r^3 f(r) \giso'(r) \D r$.
Combining this relation with Eq.~\eqref{DL_g}, we obtain a parametrization of $g(\R{})$ as a function of $\gSigma' $ and $\giso(r)$.
Expanding in powers of $\gSigma' $ the tensors
$\bm{G}_2$ and $\bm{G}_3$ defined in Eqs.~\eqref{def_G2} and \eqref{def_G3}, the traceless part of
Eq.~\eqref{eq:Sigma:v0} then reads
\begin{equation}
\label{eq:Sigmaprime}
\dot{\gSigma'} = \kappa\Einf + \Ominf \gdot \gSigma' - \gSigma' \gdot \Ominf + \left[\beta - \xi \left(\gSigma':\gSigma'\right)\right]\gSigma'
\end{equation}
where $\Einf$and $\Ominf$ are respectively the strain-rate and vorticity tensors, 
$\Einf = \left[ \nabla\bm{u}^{\infty} + (\nabla\bm{u}^{\infty})^T \right]/2$ and $
\Ominf = \left[ \nabla\bm{u}^{\infty} - (\nabla\bm{u}^{\infty})^T \right]/2$.
The coefficients $\kappa$, $\beta$ and $\xi$ are given by multiple integrals involving the isotropic pair correlation function $\giso(r)$ and its derivative (see \cite{CunyPRE21} for the detailed expressions of these coefficients).
To evaluate these coefficients, we use a simple and physically motivated parametrization of the isotropic pair correlation function $\giso(r)$, of the form
\begin{equation}
	\label{expr_alpha0}
	\giso(r)=\frac{A}{\rs}\delta(r-\rs)+H(r-\rs)
\end{equation}
with $H(x)$ the Heaviside function. The parametrization (\ref{expr_alpha0}) approximates the first shell of neighbors as a delta peak at a distance $r^*$, and considers that the pair correlation is flat for $r>r^*$. Although this form of the pair correlation is obviously an oversimplification, it already captures key features of the microstructure well above jamming with a small number of parameters.
The amplitude $A$ is fixed by assuming that the first shell of neighbors contains $6$ particles on average in two dimensions~\cite{schreck_tuning_2011}, leading to $A=3/(\pi\rho)$.
The parameter $r^*$ can be reexpressed in terms of the pressure $p=-\frac{1}{2} {\rm Tr}\gSigma$,
so that the coefficients $\kappa$, $\beta$ and $\xi$ appearing in Eq.~\eqref{eq:Sigmaprime}
are now functions of the pressure $p$ and of the average density $\rho$ (or the packing fraction $\phi=\pi\rho$).
To study the behavior of Eq.~\eqref{eq:Sigmaprime}, we thus need an equation of state for the pressure $p$.
The latter is obtained by taking the trace of Eq.~\eqref{eq:Sigma:v0}, yielding an evolution equation for $p$ (see \cite{CunyPRE21}). 
One thus obtains coupled evolution equations for $\gSigma'$ and $p$, that can be integrated numerically \cite{CunyPRE21}.
For not-too-large values of the strain rate, the pressure relaxes much faster than the deviatoric stress $\gSigma'$, and it is thus possible to use the steady-state value for $p$ in Eq.~\eqref{eq:Sigmaprime}.
This provides an equation of state for the pressure in terms of  $\phi$,  $\gSigma'$ and $\Einf$.
Here, for simplicity, we use as an approximation the isotropic equation of state $p(\phi)$, neglecting the contributions of $\Einf$ and $\gSigma'$ to the pressure.
The coefficients $\kappa(p)$, $\beta(p)$ and $\xi(p)$ in Eq.~\eqref{eq:Sigmaprime} then become constants, for a given $\phi$.
Eq.~\eqref{eq:Sigmaprime} with constant coefficients may thus be considered, at a qualitative level, as a minimal constitutive equation for the rheology of soft dense suspensions for intermediate strain rates.
It contains the minimal terms required for a yield stress fluid: a term $\propto \Einf$ to account for strain-rate dependence, the commutator $[\Ominf,\gSigma']$ as imposed by frame indifference, and a tensorial Landau-like term which generates a yield stress if $\beta>0$ and $\xi>0$.
One advantage of this formulation with respect to phenomenological approaches is that the coefficients are known in terms of microscopic parameters and $\phi$. 
There is no free parameter in our theory, yet it is able to qualitatively capture many non-trivial features of YSF. 
Furthermore, our theory can trace the origin of rheological behaviors back to the microstructure dynamics.

\begin{figure*}[t]
  \includegraphics[height=0.33\textwidth]{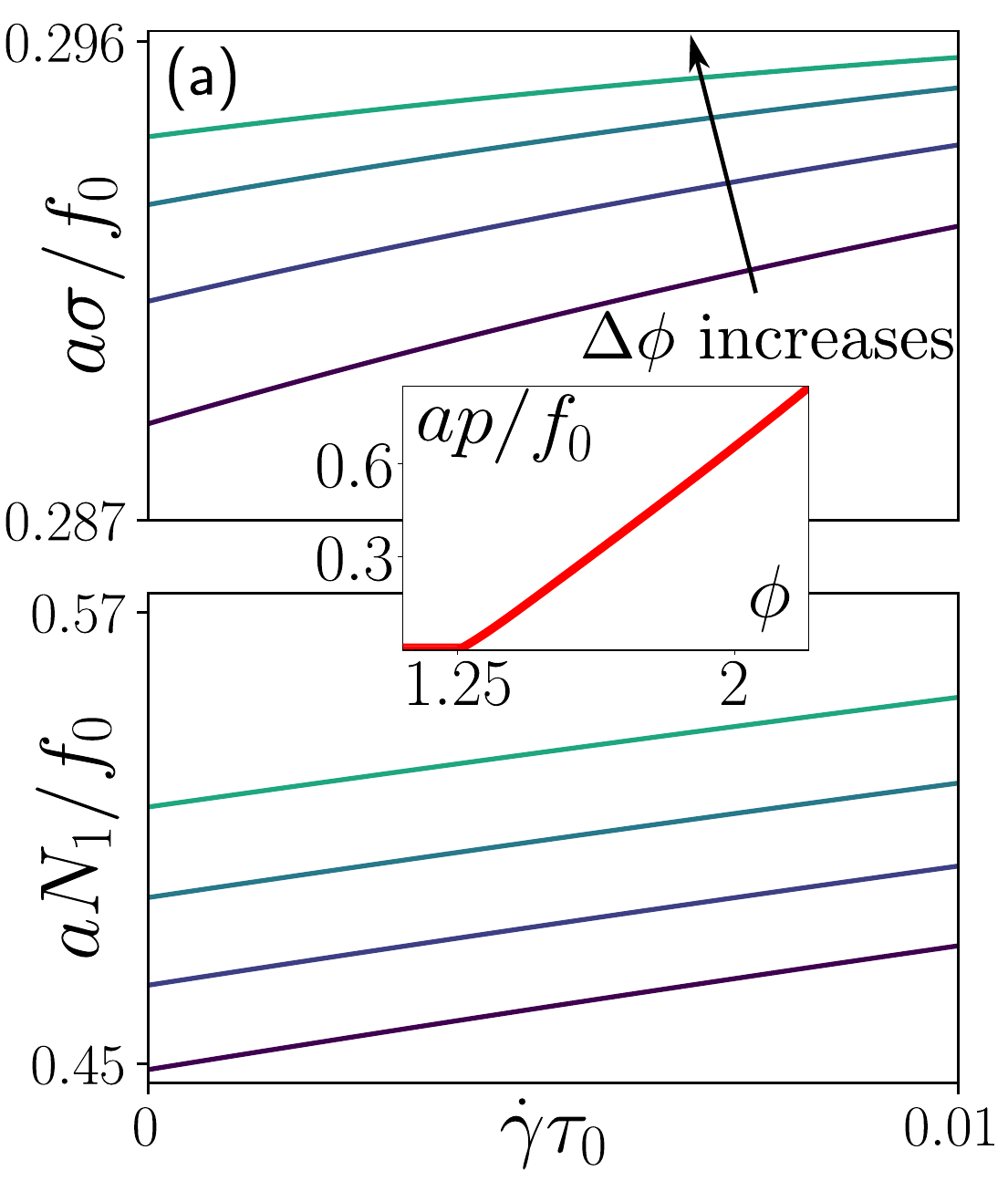}
  \hfill
  \includegraphics[height=0.33\textwidth]{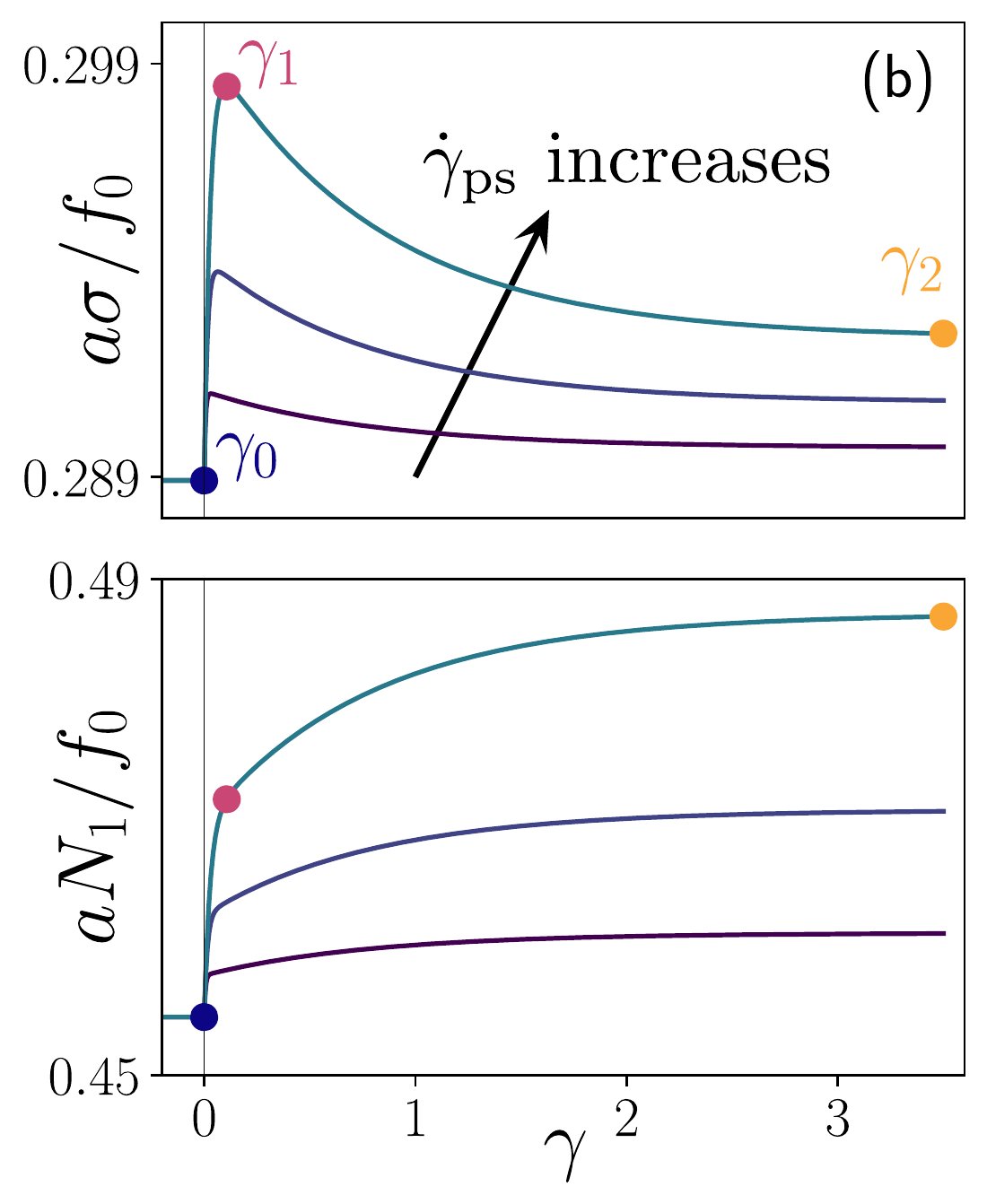}
  \hfill
  \includegraphics[height=0.33\textwidth]{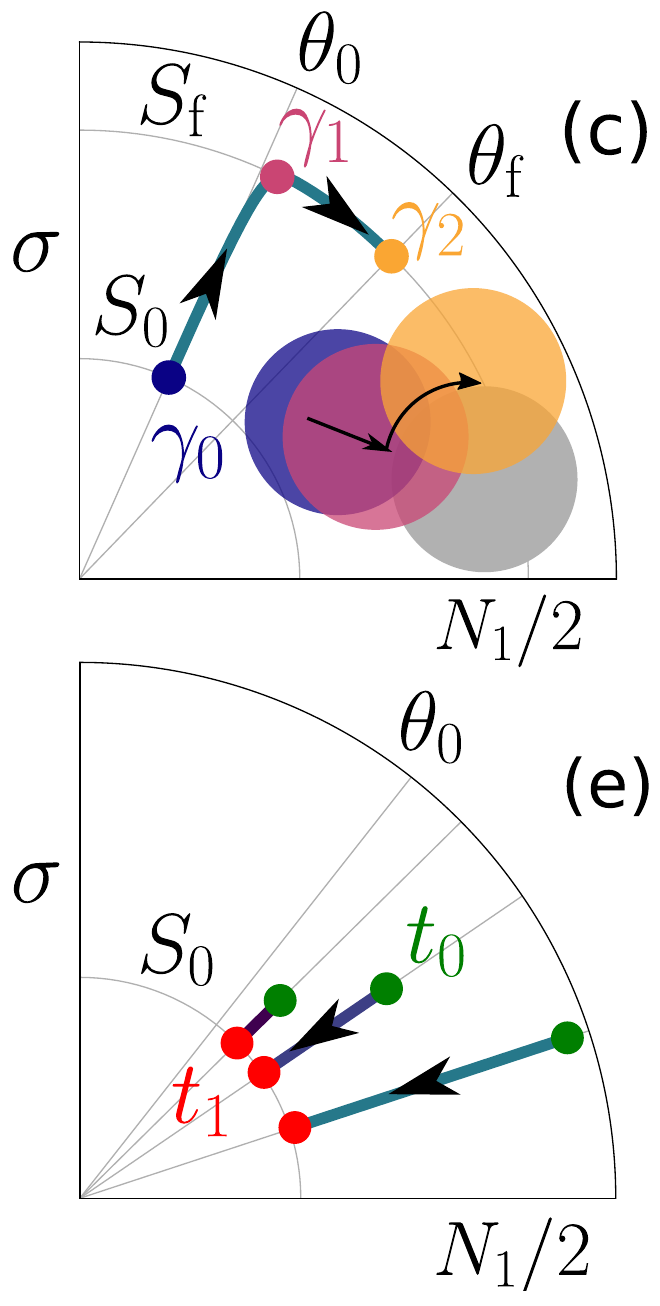}
  \hfill
  \includegraphics[height=0.33\textwidth]{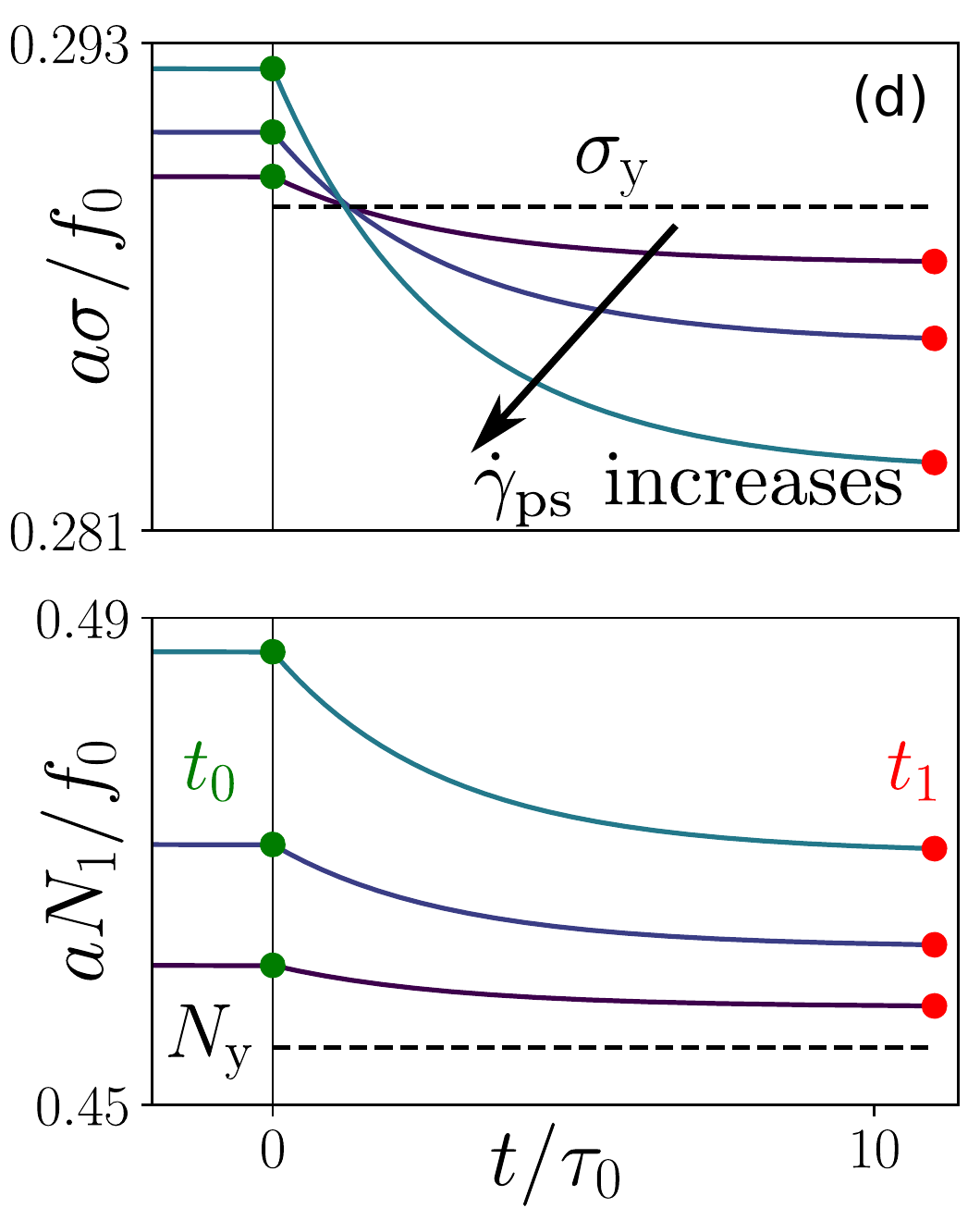}
  \caption{Particle shear stress $\sigma$ and normal stress difference $N_1$ in explicit dimensionless forms, for different protocols.
(a) Stationary flow curves for different packing fractions $\phi=\phi_{\rm J}+\Delta \phi$ with $\Delta\phi=0.01$, $0.02$, $0.03$, $0.04$ (bottom to top). Inset: pressure $p(\phi)$.
(b) Transient response at a constant shear rate $\dot{\gamma}$ after a preshear at low shear rate $\dot{\gamma}_{\rm ps}=10^{-5}$, showing an overshoot on the shear stress $\sigma$ for $\gd\tau_0 = 2\times 10^{-3}$, $5\times 10^{-3}$, $10^{-2}$ at $\phi=\phi_{\rm J}+0.01$.
(c) Schematic representation in the plane $(N_1/2,\sigma)$ of the transient response shown in panel (b), highlighting the fast dynamics of $S$ and the slow dynamics of $\theta$. Particle configurations corresponding to the colored dots indexed by $\gamma_i$ are schematically drawn with the same color code.
(d) Stress relaxation at $\dot{\gamma}=0$ after a preshear at shear rate $\dot{\gamma}_{\rm ps}\tau_0= 2\times 10^{-3}$, $5\times 10^{-3}$, $10^{-2}$, showing that a stronger preshear eventually leads to a lower value of the shear stress.
(e) Schematic representation in the plane $(N_1/2,\sigma)$ of the relaxation shown in panel (d), which occurs at constant $\theta$.
Data are obtained by numerical integration of Eq.~\eqref{eq:polar:S:theta}, using a linear force $f(r) = f_0 (r-2a)/a$ (for $r<2a$) to evaluate the coefficients.}
  \label{fig:1}
\end{figure*}

We now discuss the behavior of \eqref{eq:Sigmaprime}, using several deformation protocols.
For definiteness, we specialize to a linear repulsive contact force $f(r)=(2-r)H(2-r)$, in dimensionless form.
First, we consider the steady-state rheology under simple shear, $\bm{u}^{\infty} = \gd y \bm{e}_x$ with $x$ the flow and $y$ the gradient directions. The tensor $\gSigma'$ can be written as 
\begin{equation}
	\label{ecriture_sigma}
	\gSigma'=\Msymt{N_1/2}{\sigma},
\end{equation}
with $\sigma$ the shear stress and $N_1$ the normal stress difference.
Introducing a polar parametrization $N_1 = 2 S \cos\theta$ and $\sigma = S \sin\theta$, we get
\begin{equation}
	\label{eq:polar:S:theta}
	\left\{
	\begin{aligned}
		&\dot{S} = \frac{\kappa \gd}{2} \sin\theta + \beta S - 2\xi S^3 ,\\
		&\dot{\theta} =  \frac{\kappa \gd}{2S}  \cos\theta - \gd\,.
	\end{aligned}
	\right.
\end{equation}
Physically, $S$ is the amplitude of the stress anisotropy and $\theta$ its orientation. 
Because $\gSigma' = k \gQ$, $S$ also quantifies the deformation of the contact shell, and $\theta$ is twice its tilt angle with respect to the flow direction.

The isotropic equation of state $p(\phi)$ is such that $p(\phi)=0$ for $\phi<\phi_{\rm J}$ and $p(\phi)>0$ for $\phi>\phi_{\rm J}$, where $\phi_{\rm J} = 1.25$ is the jamming packing fraction (a value $50\%$ above the correct jamming packing fraction in two dimensions, due to the approximations made).
Numerically, we find that $p$ increases with $\phi$
[see inset of Fig.~\ref{fig:1}]. The small $\Delta\phi=\phi-\phi_{\rm J}$ behavior is linear as $p\approx 0.63\Delta\phi$, when particle simulations show a slightly larger exponent of $1.1$~\cite{vagbergPressureDistributionCritical2014}.
The coefficients $\kappa$, $\beta$ and $\xi$ are positive, leading to a yield stress fluid behavior due to the Landau-like term in the equation for $S$. The yield follows the von Mises criterion $\gSigma':\gSigma' = \beta/\xi$, as sometimes observed experimentally~\cite{ovarlezThreedimensionalJammingFlows2010} (but not always, see~\cite{de_cagny_yield_2019}).
For small $\Delta\phi$,
we find
$\kappa\approx 1.19 - 0.099\Delta\phi$,
$\beta\approx 0.16 +0.76\Delta\phi$,
and $\xi\approx 0.62 +0.0054\Delta\phi$ \cite{CunyPRE21}.
Looking perturbatively for the small $\gd$ solution, one finds
\begin{equation}
\label{eq:S:theta:perturb}
S \approx S_0+b\gd \,, \quad \theta \approx \theta_0 - \frac{\gd}{2\beta} \,,
\end{equation}
with $S_0 = \sqrt{\beta/2\xi}$, $b=(\kappa\sin\theta_0)/(4\beta)$ and $\theta_0=\cos^{-1}(2S_0/\kappa)$.
For densities higher than $\phi_{\rm m}\approx1.5$, we find that $2S_0/\kappa>1$ such that there is no stable solution. We thus consider $\phi_{\rm J} < \phi < \phi_{\rm m}$ to be the validity range of our approximations.
We therefore obtain a Bingham fluid behavior for both the (particle) shear stress component $\sigma$ 
and the normal stress difference $N_1$,
\begin{equation}
\sigma = \sigma_{\rm y} + b_{\sigma} \gd \,, \quad
N_1 = N_{\rm y} + b_N \gd \,,
\end{equation}
with yield stress values $\sigma_{\rm y}=S_0 \sin\theta_0$ and $N_{\rm y}=2 S_0 \cos\theta_0 > 0$, and prefactors
$b_{\sigma} = \kappa/(4\beta)-1/(\kappa\xi)$ and $b_N = 2(S_0/\beta)\sin\theta_0$.
Steady-state flow curves are displayed on Fig.~\ref{fig:1}(a), 
showing the increase of yield values with $\phi$.
A positive $N_1$ is also observed in experiments~\cite{de_cagny_yield_2019} and in simulations~\cite{evans_networklike_2013,liu_universality_2018}. 
Our model predicts a ratio $N_{\rm y}/\sigma_{\rm y} \approx 1.5\text{--}1.8$, 
a value slightly larger than what is found in 2D simulations of foams for which $N_{\rm y}/\sigma_{\rm y} \approx 1$~\cite{evans_networklike_2013} and much larger than the value $\approx 0.2$ observed in experiments~\cite{habibi_normal_2016,de_cagny_yield_2019} and in simulations of 3D microgels~\cite{liu_universality_2018}. 
Similarly large values of $N_{\rm y}/\sigma_{\rm y}$ are also found in tensorial versions of the SGR model~\cite{cates_tensorial_2003}.

We further explore the transient dynamics of the stress with two different protocols,  numerically integrating  Eq.~\eqref{eq:polar:S:theta}.
In the first one, we pre-shear up to steady state at a low rate $\dot{\gamma}_{\rm ps}$, and then follow the stress dynamics after suddenly increasing the shear rate to a constant value $\dot{\gamma}>\dot{\gamma}_{\rm ps}$. An overshoot is observed on the shear stress, which relaxes on an order 1 strain scale, 
in agreement with experiments~\cite{amemiya_measurement_1992,batista_colored_2006,divoux_stress_2011,younes_elusive_2020,khabazTransientDynamicsSoft2021}, while the normal stress difference increases monotonously [Fig.~\ref{fig:1}(b)], here too in agreement with numerical simulations of foams~\cite{evans_networklike_2013}. The polar representation of $\gSigma'$ gives us insights on the microscopic dynamics.
An early fast increase of $S$ corresponds to a radial compression or elongation 
of contacts at almost fixed orientation. 
After the overshoot, $S$ saturates quickly while $\theta$ starts decreasing [Fig.~\ref{fig:1}(c)]. 
This reflects a rotation of closest contacts with the vorticity, which consequently decreases (resp. increases) their shear stress (resp. normal stress difference) contribution.

We then consider the relaxation at $\dot{\gamma}=0$ after a preshear.
It has been experimentally and numerically observed that shear stress unexpectedly relaxes to a lower value for a stronger preshear \cite{mohan_microscopic_2013,mohan_build-up_2014,lidon_power-law_2017,zakhari_stress_2018}.
This effect is reproduced by our model [Fig.~\ref{fig:1}(d)], and can be understood as follows.
During the preshear phase, a higher shear rate $\gd_{\rm ps}$ leads to a smaller value $\theta_{\rm ps}$ of the angle $\theta$, according to Eq.~\eqref{eq:S:theta:perturb}.
Switching off the shear, $S$ relaxes to $S_0$ while $\theta$ keeps the value $\theta_{\rm ps}$ it had at the end of the preshear, as seen from Eq.~\eqref{eq:polar:S:theta}. One thus finds for the final values $\sigma_{\rm f}$ of the shear stress and $N_{\rm f}$ of normal stress difference
\begin{equation}
\sigma_{\rm f} = \sigma_{\rm y} - \frac{\gd_{\rm ps}}{2\kappa\xi} \,, \quad N_{\rm f} = N_{\rm y} + \frac{\sigma_{\rm y}}{\beta} \gd_{\rm ps} \,,
\end{equation}
that is, $\sigma_{\rm f}/\sigma_{\rm y}<1$ is a decreasing function of $\gd_{\rm ps}$. 
(Experiments report weaker dependencies, $\sigma_{\rm f}/\sigma_{\rm y} \propto \gd_{\rm ps}^{-0.2}$~\cite{lidon_power-law_2017} or even $\sigma_{\rm f}/\sigma_{\rm y} \propto -\log \gd_{\rm ps}$~\cite{mohan_build-up_2014}.) In contrast,
$N_{\rm f} > N_{\rm y}$ and increases with $\gd_{\rm ps}$. The opposite trends on $N_{\rm f}$ and $\sigma_{\rm f}$ are a consequence of the von Mises yield criterion our model predicts.
Interestingly, the relaxation curves for different preshear values approximately intersect at a same time $t^*$, and for shear stress values close to the yield stress. Those predictions on the shear stress relaxation are consistent with experimental observations \cite{mohan_microscopic_2013}.
The predicted relaxation time scale ($\approx 3\tau_0$) is however much smaller than measured in~\cite{mohan_microscopic_2013}, but consistent with~\cite{zakhari_stress_2018}.
The relaxation dynamics is well understood in the polar coordinates $(S,\theta)$, as the dynamics occurs at a constant value $\theta_{\rm ps}$ determined by the preshear, until $S$ has relaxed to $S_0$ [Fig.~\ref{fig:1}(e)].
As $\theta_{\rm ps}$ decreases with the preshear $\gd_{\rm ps}$, a stronger preshear leads to a lower final value of $\sigma$.
Linearizing the dynamics \eqref{eq:polar:S:theta} of $S$ around $S_0$, one finds that curves for different $\gd_{\rm ps}$ cross precisely at $\sigma=\sigma_{\rm y}$, at a (dimensionless) time
$t^*=(2\beta)^{-1}\ln[\xi\kappa^2/(2\beta)-1]$.

To sum up, we have derived from the microscopic dynamics a minimal tensorial constitutive model for jammed soft suspensions in two dimensions. This constitutive model accounts at a qualitative level for a number of non-trivial features, like the  existence of yield stresses for both the shear stress and the normal stress difference, a stress overshoot on step change of shear rate, or the preshear dependence of residual stresses during stress relaxation.
One of the main interests of the present result with respect to more standard phenomenological approaches is that all parameters are fixed by
the particle-level dynamics, thus
shedding light on the microscopic mechanisms responsible for different types of macroscopic rheological behaviors.
Moreover our approach links macroscopic phenomena like overshoots in transient response to the particle-level behavior, which could stimulate further numerical or experimental work in this direction.
Note that our approach, which partly neglects correlations, is valid sufficiently far away from both the jamming density and the yielding transition (i.e., when $\phi-\phi_{\rm J}$ and $\dot\gamma$ are not too small), so that static and dynamic critical fluctuations do not dominate the physics.
In contrast to MCT \cite{fuchs_theory_2002,brader_glass_2009}, our theory directly addresses the dynamics of athermal, out-of-equilibrium systems without the need for a nearby reference equilibrium state, and is based on a physically more transparent real-space description.

Future work will investigate the role of the precise form of the repulsive force, and extend the present approach to dense soft suspensions just below jamming, as well as to the potentially richer three-dimensional case.
On a longer term, it would be useful to improve the approximations made
(in particular the Kirkwood closure and the parametrization of the pair correlation function)
so as to describe more accurately the low-shear-rate behavior of jammed soft suspensions, when localized plastic events dominate the relaxation processes.
Such plastic events have a four-fold symmetry which is not captured by the parametrization \eqref{DL_g} of the pair correlation function. Generalizing the present approach by including a fourth order harmonic in $g(\mathbf{r})$
might grant access to the HB rheology.

\paragraph{Acknowledgments}
This work is supported by the French National Research Agency in the framework of the "Investissements d'avenir" program (ANR-15-IDEX-02).

\bibliographystyle{apsrev4-2}
\bibliography{biblio}

\end{document}